\documentclass[twoside,twocolumn]{article}
\usepackage{geometry}
\usepackage{setspace}

\usepackage{hyperref}
\usepackage{amsmath}
\usepackage{amssymb}
\usepackage{authblk}
\usepackage{float}
\usepackage{graphicx}
\usepackage[version=3]{mhchem}
\usepackage{bm}

\usepackage[backend=biber, style=numeric-comp, sorting=none]{biblatex}
\title{The tunneling splitting and the Kramers theory of activated processes}
\author[1]{Pierpaolo Pravatto}
\author[1]{Barbara Fresch}
\author[1]{Giorgio J. Moro}
\affil[1]{Università degli studi di Padova, Dipartimento di Scienze Chimiche, Via Marzolo 1 - 35131 Padova (Italy)}

\begin{document}

\twocolumn[
    \begin{@twocolumnfalse}
        \maketitle
        \begin{abstract}
            The study of tunneling splitting is fundamental to get insight into the dynamics of a multitude of molecular systems. In this paper, a novel approach to the analysis of the ground-state tunneling splitting is presented and explicitly applied to one-dimensional systems. The isomorphism between the Fokker-Planck-Smoluchowski operator and the Born-Oppenheimer quantum Hamiltonian is the key element of this method. The localization function approach, used in the field of stochastic processes to study the Kramers problem, leads to a simple, yet asymptotically justified, integral approximation for the tunneling splitting. The comparison with exact values of the tunneling splittings shows a much better accuracy than WKB semiclassical estimates.\\
        \end{abstract}
    \end{@twocolumnfalse}
]

\section{Introduction}
    In the study of symmetrical quantum systems characterized by two or more identical potential wells, the term tunneling splitting denotes the energy separation arising from the mixing of overlapping degenerate states localized at different potential wells. These splitting are observed by spectroscopical means in many molecular systems ranging from simple molecules, such as ammonia~\cite{Good1946}, malondialdehyde~\cite{Firth1991, Baba1999} and tropolone~\cite{Tanaka1999}, to molecular clusters~\cite{Keutsch2001}.\\
    
    The study of tunneling splittings plays a fundamental role in understanding the dynamics of these molecular systems and their theoretical evaluation is still an active field of research. Over the years several methods have been developed on the basis of the quantum description of the nuclear motion in molecules. Basically, they can be classified into two categories: 1) computational approaches aiming at solving exactly the underlying mathematical problem and 2) approximate methods holding in the semi-classical limit. The numerical diagonalization of the nuclear Hamiltonian~\cite{Chen1998, Wang2018, Althorpe1995}, path integral based simulations~\cite{Matyus2016} and Diffusion Monte Carlo~\cite{Gregory1995} are some notable examples of the former approach while WKB methods~\cite{LandauQM, Garg}  and instanton theory~\cite{Richardson2011}  are the most widely applied approximations.\\
    
    In this work, we introduce a different methodology based on the well-known isomorphism~\cite{elber2020, risken1989} between the quantum Hamiltonian and the time evolution operator of the Fokker-Planck-Smoluchowski (FPS) equation that describes diffusion motion in the over-damped limit~\cite{Gardiner}. In the presence of large enough barriers between stable configurations, the FPS equation represents the simplest model for describing activated processes whose kinetic rates are determined by the low-lying eigenvalues of the FPS evolution operator. Starting from H. Kramers in 1940~\cite{Kramers1940}, this has been a very active field of research, sometimes referred to as Kramers theory, which has produced a variety of tools for the description of activated processes and for the evaluation of the corresponding rate constants~\cite{Borkovec}. Because of the isomorphism with the quantum Hamiltonian underlying such a stochastic description, the rate constants of activated processes correspond to tunneling splittings. Then one might consider the use of the methods of Kramers theory for evaluating the low-lying eigenvalues of FPS evolution operator in order to calculate the tunneling splittings.\\  
    
    In order to explore the feasibility and the benefits of such a procedure, as well as its limits, we shall consider here one-dimensional model systems. Furthermore, we shall employ the site localization functions~\cite{Moro1995, MoroNordio, Moro1986} in order to evaluate the first non-vanishing eigenvalue of the FPS evolution operator corresponding to the tunneling splitting. For significant barriers, this method provides a rather accurate approximation according to an integral form which allows a direct analysis of the tunneling splitting in terms of the features of the quantum potential.\\ 
    
    In the next section, the isomorphism between the quantum Hamiltonian and the FPS evolution operator is discussed for the case of bi-stable one-dimensional systems, in order to formalize the connection between the kinetic rate constant and the tunneling splitting. Then the integral form for the tunneling splitting is derived by means of the site localization functions. In section~\ref{sec:models} we apply this method to specific cases to illustrate the typical results and the accuracy of the calculated tunneling splitting. In the same section, we examine a crucial component of our method: the modeling of the equilibrium distribution of the FPS equation which ensures the correspondence with a proper quantum potential. We propose a convenient solution in terms of the linear combination of two Gaussian distributions. The final section reports the main conclusions of our analysis together with a  discussion of the challenges posed by the extension of the proposed method to multi-dimensional tunneling problems.

\section{Theory}\label{sec:theory}
    One-dimensional systems provide the most convenient framework for examining the connections existing between a quantum property like the tunneling splitting and the classical stochastic picture of activated processes according to the Kramers theory. Given a suitable bi-stable symmetric potential $V(x)=V(-x)$ dependent on the real-valued physical coordinate $x$  (denoted as quantum potential in the following)  and the standard form of the quantum Hamiltonian,
    \begin{equation}\label{BO-Hamiltonian}
        \hat{H}=-\frac{\hbar^2}{2m}\frac{\partial^2}{\partial x^2} + V(x),
    \end{equation}
    the time independent Schrödinger equation, $\hat{H}\psi_n(x)=E_n\psi_n(x)$ for $n=0, 1, 2,...$, is invoked to identify the eigenstates $\psi_n(x)$ and the corresponding eigenvalues $E_n$ ordered in magnitude as $E_n < E_{n+1}$. As long as the potential barrier at the origin is finite, degeneracy is excluded and the eigenstates $\psi_n(x)$ can be specified as real functions. The tunneling through the barrier determines the mixing of states localized about the potential minima~\cite{LandauQM}, resulting in a ground state energy separation which is quantified by the tunneling splitting:
    \begin{equation}
        \delta E_1:=E_1-E_0
    \end{equation}
    The tunneling splitting is a measure of the coupling between the states localized in the potential wells, and, in general, it decreases in magnitude when the potential energy barrier is increased. In what follows we will suppose that the barrier is large enough to determine a significant spectral gap between the first two eigenstates and the others: $E_1-E_0 \ll E_2-E_0$. In this framework a fundamental issue arises: what is the relation between the shape of the quantum potential $V(x)$ and the tunneling splitting?\\
    
    In order to provide an answer to such a question, one can solve the problem with standard methods of quantum mechanics like WKB theory or the numerical diagonalization of the Hamiltonian matrix representation. In this work, we intend to explore an alternative procedure based on the classical stochastic description of activated processes according to the Fokker-Planck-Smoluchowski (FPS) equation. This equation describes the time-evolution of the probability distribution $\rho(x,t)$ on the configurational coordinate $x$~\cite{Gardiner} and it is completely specified by the diffusion coefficient $D$ and the equilibrium distribution $\rho_{eq}(x)$. The equilibrium distribution is often specified according to the Boltzmann canonical distribution  $\rho_{eq}(x)\propto \exp{\left[-U(x)\right]}$ determined by the mean-field potential $U(x)$, here expressed in $k_BT$ units. A constant diffusion coefficient is considered since a coordinate dependent form $D(x)$ would be relevant only for the quantum Hamiltonian with a coordinate dependent effective mass $m(x)$, a case that is not examined here for the sake of simplicity. The FPS time-evolution can be specified in the symmetrized form, denoted also as the Schroedinger representation~\cite{elber2020}
    according to the equation:
    \begin{equation}\label{FPS_evolution}
        \frac{\partial\Tilde{\rho}(x,t)}{\partial t}=-\hat{\Gamma}\Tilde{\rho}(x,t)
    \end{equation}
    where $\Tilde{\rho}(x,t):=\rho_{eq}(x)^{-\frac{1}{2}}\rho(x,t)$ represents a modified distribution and the corresponding symmetrized FPS operator $\hat{\Gamma}$ is written as:
    \begin{equation}\label{symm_FPS_operator}
        \begin{split}
            \hat{\Gamma}&:=-D\rho_{eq}(x)^{-\frac{1}{2}}\frac{\partial}{\partial x}\rho_{eq}(x)\frac{\partial}{\partial x}\rho_{eq}(x)^{-\frac{1}{2}}=\\
            &=-D\frac{\partial^2}{\partial x^2} + D\rho_{eq}(x)^{-\frac{1}{2}}\frac{\partial^2 \rho_{eq}(x)^{\frac{1}{2}}}{\partial x^2}.
        \end{split}
    \end{equation}
    Notice that, for bounded equilibrium distributions $\rho_{eq}(x)$, $\hat{\Gamma}$ is a positive semi-definite operator, that is $\langle f\rho_{eq}^{1/2} \rvert \hat{\Gamma} \lvert f\rho_{eq}^{1/2}\rangle\geq0 \; \forall f(x)$, having $\rho_{eq}(x)^{1/2}$ as the eigenfunction with a vanishing eigenvalue, $\hat{\Gamma}\rho_{eq}(x)^{1/2}=0$, in correspondence of the stationary solution of the FPS evolution from eq.~\ref{FPS_evolution}. \\
    
    Considering that both the Hamiltonian from eq.~\ref{BO-Hamiltonian} and the FPS evolution operator in eq.~\ref{symm_FPS_operator} are second-order differential operators, we wonder whether in some circumstances they become equivalent. This would require a  positive semi-definite Hamiltonian with a vanishing eigenvalue. This condition is verified if we use the ground state energy $E_0$ as the origin of the energy scale for the quantum system, that is if we consider the shifted Hamiltonian:
    \begin{equation}\label{shifted_Hamitonian}
        \delta\hat{H}:=\hat{H}-E_0=-\frac{\hbar^2}{2m}\frac{\partial^2}{\partial x^2} + \delta V(x),
    \end{equation}
    where $\delta V(x)$ is the corresponding shifted quantum potential that, according to time independent Schrödinger equation for the ground state, can be specified as:
    \begin{equation}\label{shifted_quantum_potential}
        \delta V(x):=V(x)-E_0=\frac{\hbar^2}{2m}\psi_0(x)^{-1}\frac{\partial^2 \psi_0(x)}{\partial x^2}.
    \end{equation}
    Then, by comparing eq.~\ref{shifted_Hamitonian} with eq.~\ref{symm_FPS_operator}, we can conclude that the shifted Hamiltonian becomes proportional to the FPS evolution operator:
    \begin{equation}\label{equivalence_FPS_Hamiltonian}
        \delta\hat{H}=\hbar\hat{\Gamma}
    \end{equation}
    if the particle mass $m$ and the ground state $\psi_0(x)$ are assigned, on the basis of the ingredients of the Fokker-Planck-Smoluchowski operator, as follows:
    \begin{align}
        m&=\frac{\hbar}{2D} \label{mass_equivalence}\\
        \psi_0(x)&=\rho_{eq}(x)^{\frac{1}{2}} \label{ground-state_equivalence}
    \end{align}
    The proportionality factor $\hbar$ in eq.~\ref{equivalence_FPS_Hamiltonian} is required to assure the dimensionality congruence between the Hamiltonian (energy) and the FPS evolution operator (inverse of time). Thus, once the FPS evolution operator is selected, the derivation of its quantum counterpart $\delta\hat{H}$ is straightforward.\\

    It should be stressed that the other way around is not trivial at all since, once the Hamiltonian in eq.~\ref{BO-Hamiltonian} is specified according to the particle mass $m$ and the quantum potential $V(x)$, one has to solve the eigenvalue problem $\hat{H}\psi_0=E_0\psi_0$ for the ground state in order to find the equilibrium distribution $\rho_{eq}(x)=\psi_0(x)^2$. In other words, in spite of the equivalence specified by eq~\ref{equivalence_FPS_Hamiltonian}, a procedural asymmetry holds between the quantum problem and the stochastic model: the shifted Hamiltonian is obtained by default once the FPS evolution operator is supplied, while the derivation of the latter from a given Hamiltonian (eq.~\ref{BO-Hamiltonian}) is not a simple task.
    
    On the other hand, the equivalence specified by eq.~\ref{equivalence_FPS_Hamiltonian} implies that the quantum eigenstates can be recovered as the eigenfunctions of the FPS evolution operator:
    \begin{equation}\label{FPS_Hamiltonian_eigenproblem}
        \hat{\Gamma}\psi_n(x)=\frac{\delta E_n}{\hbar} \psi_n(x)
    \end{equation}
    where $\delta E_n:=E_n-E_0$ are the shifted Hamiltonian eigenvalues. We stress that, from the point of view of eq.~\ref{FPS_Hamiltonian_eigenproblem}, the ground eigenstate $\psi_0(x)$ is not an unknown element since it is determined by the square root of the equilibrium distribution, which enters in the definition of the stochastic operator $\hat{\Gamma}$.\\
    
    The previous analysis of the connection between the quantum and the stochastic representations is quite general and does not refer to a particular system. Let us now apply it to the specific case of the tunneling splitting problem when the ground state is symmetric, $\psi_0(x)=\psi_0(-x)$, and described by two local distributions symmetrically arranged about the origin at $x=0$. According to eq.~\ref{ground-state_equivalence}, the same behavior has to be attributed to the equilibrium distribution $\rho_{eq}(x)$ and this corresponds to a bi-stable mean-field potential $U(x)$ with two equivalent minima separated by a barrier having its maximum located at the origin. This is the natural framework for the stochastic description of the interconversion between two chemical species that is controlled by an energy barrier crossing. The macroscopic picture is provided by the kinetic mechanism $A \overset{k}{\rightarrow} B, \, B \overset{k}{\rightarrow} A $ for chemical species $A$ and $B$ characterized by the same equilibrium concentration and, therefore, by the same unimolecular rate constant $k$. The rate constant controls the relaxation to the equilibrium of the concentrations according to the exponential decay $\exp{(-2kt)}$.\\
    
    An equivalent exponential relaxation to equilibrium is recovered from the solution of the Fokker-Planck-Smoluchowski equation if the energy barrier is large enough to determine a significant gap between the first non-vanishing eigenvalue $\delta E_1/\hbar$ and the other ones, that is if $\delta E_1 \ll \delta E_2$. In such a case the long-time decay of non-equilibrium distributions is controlled by $\exp{(-\delta E_1t/\hbar)}$. Comparing this result with the macroscopic kinetics of the interconversion $\ce{A <=> B}$, a direct relation $k=\delta E_1/2\hbar$ between the rate constant $k$ and the relaxation described by the FPS evolution operator is obtained. Starting from the milestone contribution by Kramers work~\cite{Kramers1940}, several methods have been elaborated in order to describe the rate of activated processes on the basis of a stochastic description of the motion~\cite{Borkovec, elber2020, risken1989}. In the case of Fokker-Planck-Smoluchowski models like eq.~\ref{symm_FPS_operator}, Kramers-type analyses can be interpreted as procedures for the estimate of the first non-vanishing eigenvalue $\delta E_1/\hbar$. Let us keep in mind that, because of the equivalence with the quantum problem according to eq.~\ref{equivalence_FPS_Hamiltonian}, estimating $\delta E_1$ means to estimate the tunneling splitting of the corresponding quantum problem. \\
    
    In the presence of large enough barriers we shall employ a procedure based on the localization function~\cite{Moro1995, MoroNordio} for the calculation of $\delta E_1/\hbar$ by exploiting the condition on the eigenvalue gap $\delta E_1/\hbar \ll \delta E_n/\hbar$ for $n \geq 2$. This implies that, in the scale of typical $\hat{\Gamma}$ eigenvalues, $\delta E_1/\hbar$ is negligible and the corresponding eigenfunction $\psi_1(x)$ can be estimated from the approximation
    \begin{equation}\label{FPS_asymptotic_approx}
        \hat{\Gamma}\psi_1(x)=\frac{\delta E_1}{\hbar} \psi_1(x)\simeq 0
    \end{equation}
	 that leads to the equation
	 \begin{equation}\label{FPS_asymptotic_approx_2}
	     \rho_{eq}(x)\frac{\partial g(x)}{\partial x} = \text{constant},
	 \end{equation}
	 where we have introduced the so-called localization function $g(x):=\psi_1(x)/\rho_{eq}(x)^{1/2}$ displaying odd parity, $g(-x)=-g(x)$.\\
	 
	 As a matter of fact, the full solution of the approximation in eq.~\ref{FPS_asymptotic_approx} does not provide a function with a finite norm. Still, a good approximation to $\psi_1(x)$ is recovered by matching the $g(x)=\text{constant}$ solution of eq.~\ref{FPS_asymptotic_approx_2} outside the domain between the two potential minima, with the $x$-dependent solution of the same equation between the two minima:
	 \begin{equation}\label{localization_function}
	    g(x):=
	    \begin{cases}
            -1 & x \leq -x_m \\
            I^{-1}\int_0^{x} \frac{1}{\rho_{eq}(y)}dy & -x_m \leq x \leq x_m \\
            1 & x \geq x_m
        \end{cases}
    \end{equation}
    where $\pm x_m$ are the locations of the minima of the mean-field potential and the constant $I$ denotes the integral:
    \begin{equation}\label{I_integral}
        I:=\int_0^{x_m} \frac{1}{\rho_{eq}(y)}dy
    \end{equation}
    Notice that for equilibrium distributions localized about the potential minima, the change on the values of the $g(x)$ function is confined to a narrow domain centered at the saddle point of $U(x)$~\cite{MoroNordio, Moro1995, Moro1986}. This motivates the name localization function since the linear combinations $\left[1\pm g(x)\right]/2$ resemble unitary step functions which select the domain of attraction~\cite{Schuss} of the two stable states (i.e., the minima of the mean-field potential) at the two sides of the potential maximum. The functions $\left[1\pm g(x)\right]/2$, when multiplying a generic distribution, have the effect of localizing the distribution in the corresponding domain of attraction. It should also be mentioned that functions $\left[1\pm g(x)\right]/2$ can be identified with the so-called committor functions~\cite{E2010, Berezhkovskii2019, Berezhkovskii2021} which represent a fundamental tool for the analysis of transition paths of activated processes. \\

    The introduction of the localization function given in eq.~\ref{localization_function} leads to an explicit (yet approximated) form of the eigenfunction  $\psi_1(x)=g(x)\rho_{eq}(x)^{1/2}$. Therefore one can evaluate the corresponding eigenvalue from the expectation value of FPS evolution operator as:
    \begin{equation}\label{expr_tunneling_splitting}
        \frac{\delta E_1^g}{\hbar}:=\frac{\langle g\,\rho_{eq}^{1/2}\lvert \hat{\Gamma} \rvert g\,\rho_{eq}^{1/2} \rangle}{\langle g\,\rho_{eq}^{1/2}\rvert g\,\rho_{eq}^{1/2} \rangle} = \frac{2D}{I\langle g \rvert \rho_{eq} \lvert g \rangle},
    \end{equation}
    where the integration by part has been used to obtain the r.h.s. of this equation. The notation $\delta E_1^g$ has been introduced to stress that this is an approximation of the tunneling splitting based on the localization function $g$. We believe that eq.~\ref{expr_tunneling_splitting} is an important and useful result of the stochastic analysis, since it provides an explicit relation for the tunneling splitting in the form of a rather accurate approximation as shown in the next section. Of course it requires the numerical calculation of the integrals for the constant $I$ and for the normalization coefficient $\langle g \rvert \rho_{eq} \lvert g \rangle$ which, however, is very close to unity because of the step-like behavior of the localization function.  It should be stressed that the symmetry of FPS evolution operator, or of the equivalent Hamiltonian $\delta \hat{H}$, allows one to be more precise about the nature of the approximation. Indeed, $g(x)\rho_{eq}(x)^{1/2}$ is orthogonal to the ground state $\psi_0(x)$ and, therefore, according to the Rayleigh-Ritz theorem the approximation $\delta E_1^g$ necessarily overestimates the tunneling splitting, $\delta E_1 < \delta E_1^g$. This opens the possibility of a further optimization, by means of variational procedures, of the localization function and of the tunneling splitting $\delta E_1^g$ that will not be examined here.\\
    
    At this point however, the relation of the tunneling splitting with the quantum potential, which is main issue in the applications to molecular systems, is still missing. This, in principle, can be recovered by taking into account that the shifted quantum potential and the mean-field potential are related as:
    \begin{equation}\label{quantum_vs_thermodynamic_potentials}
        \delta V(x)=\frac{\hbar^2}{2m}\left[\frac{U'(x)^2}{4}-\frac{U''(x)}{2}\right]
    \end{equation}
    as one derives form eq.~\ref{shifted_quantum_potential} by specifying the ground eigenstate as $\psi_0(x)=\rho_{eq}(x)^{1/2}\propto \exp{\left[-U(x)/2\right]}$.\\
    
	In conclusion, the equivalence between the quantum Hamiltonian and the Fokker-Planck-Smoluchowski evolution operator allows an alternative procedure for the analysis of the tunneling splitting. The starting point is the formulation of the Fokker-Planck-Smoluchowski evolution operator in terms of a well-defined model for the equilibrium distribution $\rho_{eq}(x)$ or, equivalently, for the mean-field potential $U(x)$. The quantum potential can then be immediately obtained by means of eq.~\ref{quantum_vs_thermodynamic_potentials}. Of course, the mean-field potential $U(x)$ has to be properly chosen in order to recover a quantum potential with the desired shape. By resorting to the theory of activated processes, one can easily evaluate the tunneling splitting according to the integral approximation from eq.~\ref{expr_tunneling_splitting}. In the next section, we shall examine how this procedure can be implemented, the difficulties to be overcome and what are the most important results.

\section{Analysis of model systems}\label{sec:models}
    The implementation of the procedure described in the previous section requires, as the first step, the definition of the Fokker-Planck-Smoluchowski model according to the equilibrium distribution or the mean-field potential. The mean-field potential $U(x)$ for a specific molecular process is usually unknown. However, essential knowledge on the activated dynamics can be acquired by introducing a parametrized form of the mean-field potential whose barrier height and shape can be controlled.\\
    
\begin{figure}[ht]
        \centering
        \includegraphics[width=\columnwidth]{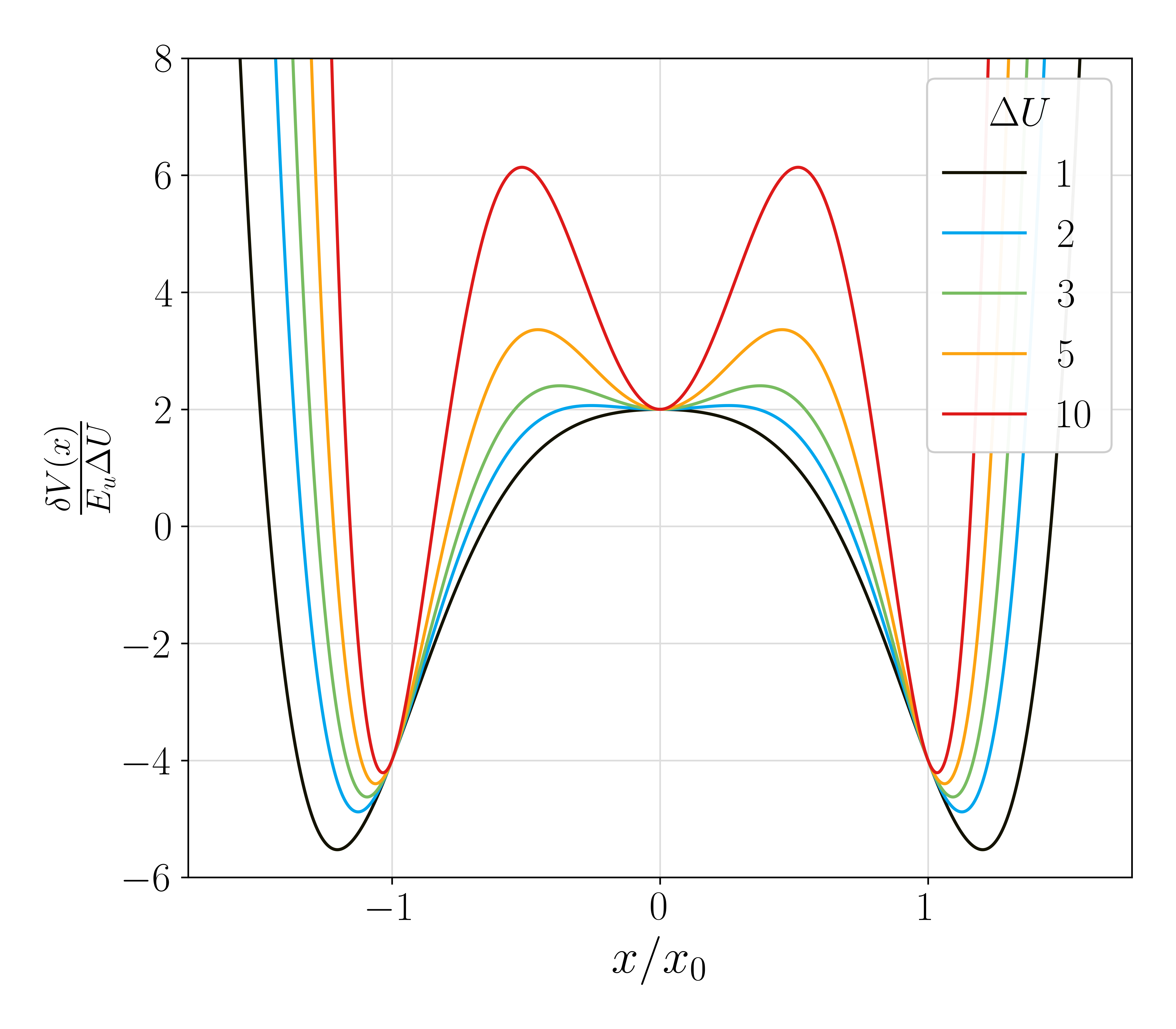}
        \caption{Quantum potential $\delta V(x)$ resulting from the quartic mean-field potential $U(x)=\Delta U (1-x^2/x_0^2)^2$ for different values (reported in the figure) of the barrier height $\Delta U$. The quantum potential is evaluated in $E_u$ units and is scaled by the barrier height $\Delta U$ in order to get roughly superimposable profiles.}
        \label{fig:Fig_1}
    \end{figure}

    We should implement this approach, keeping in mind that our objective is to select the mean-field potential which generates the quantum potential of interest according to eq.~\ref{quantum_vs_thermodynamic_potentials}. The correspondence between the mean-field potential of the stochastic problem and the quantum potential is however not always straightforward and, often, models routinely adopted in the field of activated processes do not produce well-behaving double minimum profiles for the quantum potential. Let us consider as an example the mean-field potential specified as $U(x)=s(x) \Delta U$ where $\Delta U$ sets the barrier height (in $k_B T$ units) and $s(x)=s(-x)$ represents a parameterized "shape function", responding to the condition $s(0)-s(\pm x_m )=1$, with two minima at $x=\pm x_m$ and a saddle point at the origin. A simple example is the quartic polynomial $s(x)=\left(1-x^2/x_0^2\right)^2$ where $x_0$ is the length scale determined by the minimum location. In figure~\ref{fig:Fig_1} we have drawn the shape of the quantum potential $\delta V(x)$ from eq.~\ref{quantum_vs_thermodynamic_potentials} resulting from this mean-field potential for different values of the barrier height $\Delta U$. While for low values of the barrier height, $\Delta U \leq 1$, the typical profile of a bi-stable quantum potential is found, for increasing barrier heights an additional minimum emerges at the origin. Such behavior is rather general and it is recovered independently of the specific model employed for the shape function. Indeed, by considering the large barrier limit of eq.~\ref{quantum_vs_thermodynamic_potentials} we get, $\delta V(x)=\Delta U^2(\hbar^2/8m)s'(x)^2$ as the leading contribution in $\Delta U$. Correspondingly minima of the quantum potential are found at the extremal points of the shape function where $s'(x)=0$, that is not only at the minima of $s(x)$ but also at its local maximum for $x=0$. In other words, by increasing the barrier height $\Delta U$ of the mean-field potential for a given shape function $s(x)$, one produces a change on the quantum potential from a standard bi-stable form to a completely different profile with three minima.\\
    
    In order to report in dimensionless form the quantum potential in fig.~\ref{fig:Fig_1} and any other energy parameter in what follows, we employ the energy unit:
    \begin{equation}\label{energy_unit}
        E_u:=\frac{\hbar^2}{2mx_0^2}
    \end{equation}
    A typical $E_u$ value of $0.8 \text{kJ/mol}$ in molar units or $67 \text{cm}^{-1}$ wavenumbers is recovered for $1$\r{A} displacement of an hydrogen atom between the two minima at $\pm x_0$.\\
    
    The example of the quartic potential points out that we need to pay special attention in designing a parameterized form of the FPS model in order to be compatible with a bistable quantum potential. We have found that the direct modeling of the equilibrium distribution by means of Gaussian functions can provide an effective solution to the problem. In the simplest formulation, inspired by the picture of the quantum ground state deriving from two distinct localized contributions, the equilibrium distribution is parameterized as the linear combination of two symmetrically arranged Gaussian profiles,
    \begin{equation}\label{rho_simple_gaussian}
        \rho_{eq}(x):=\frac{e^{-\frac{(x-x_0)^2}{2\sigma^2}}+e^{-\frac{(x+x_0)^2}{2\sigma^2}}}{\sqrt{8\pi\sigma^2}}
    \end{equation}
    with two independent parameters: the width $\sigma$ of the Gaussians and the center $x_0$ of the rightmost one. The corresponding mean-field potential can be specified as:
    \begin{equation}\label{U_simple_gaussian}
        \begin{split}
            U(x)&=-\ln\left[\sqrt{8\pi\sigma^2}\rho_{eq}(x)\right]=\\
            &=\frac{x_0^2+x^2}{2\sigma^2}-\ln{\left( e^{\frac{xx_0}{\sigma^2}}+e^{-\frac{xx_0}{\sigma^2}}\right)}
        \end{split}
    \end{equation}
    where we have introduced a suitable factor multiplying the equilibrium distribution in order to deal with an adimensional argument of the logarithm. According to eq.~\ref{quantum_vs_thermodynamic_potentials}, the corresponding quantum potential is given as:
    \begin{equation}\label{V_simple_gaussian}
        \begin{split}
            \frac{\delta V(x)}{E_u}=\frac{x_0^4}{4\sigma ^4}\left[ \frac{x}{x_0}-\tanh{\left(\frac{xx_0}{\sigma^2} \right) }\right]^2+\\
            +\frac{x_0^4}{2\sigma ^4} \text{sech}^2\left(\frac{xx_0}{\sigma^2}\right)-\frac{x_0^2}{2\sigma ^2}
        \end{split}
    \end{equation}
    where the standard notation for hyperbolic functions has been employed.\\
    
\begin{figure}[ht]
        \centering
        \includegraphics[width=0.9\columnwidth]{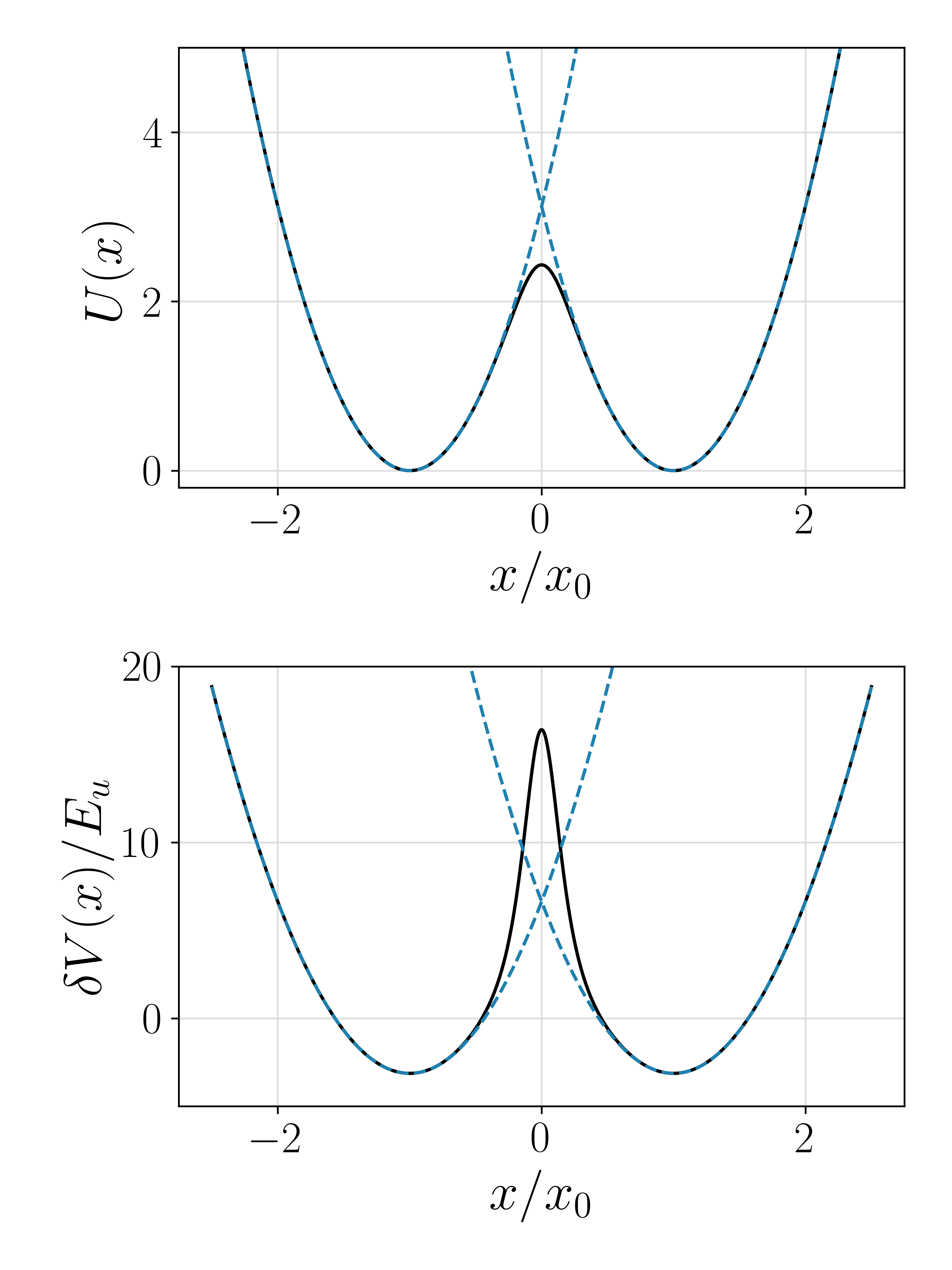}
        \caption{Mean-field potential eq.~\ref{U_simple_gaussian} (upper panel) and quantum potential eq.~\ref{V_simple_gaussian} (lower panel) derived from equilibrium distribution of eq.~\ref{rho_simple_gaussian} for $\sigma/x_0=0.4$. The dashed lines represent their parabolic approximations eq.~\ref{U_min_simple} and eq.~\ref{V_min_simple}.}
        \label{fig:Fig_2}
    \end{figure}

    The maxima of $\rho_{eq}(x)$, and the minima of $U(x)$ as well, are located near the centers of the Gaussian distributions, $x_m\simeq x_0$, with displacements, in units of $x_0$, of the order of $S:=\exp{\left[-2x_0^2/\sigma^2\right]}$. The ratio $\sigma/x_0$ plays the role of the control parameter of the model. We can employ the parameter $S$ to quantify the superposition between the Gaussians, with the two limits of a vanishing superposition ($S=0$) for infinitely narrow Gaussians, $\sigma/x_0 \rightarrow 0$, and of the full superposition ($S=1$) for a vanishing separation, $x_0/\sigma \rightarrow 0$. The objective of describing the quantum tunneling requires that the two local Gaussian contributions should be well separated and this, in turn, calls for a small superposition coefficient $S$. In the following, we assume that $S$ is small enough to allow the identification of the mean-field minima with the centers of the Gaussians, $x_m=x_0$. In the reported calculations we have considered, for the control parameter, the range $0<\sigma/x_0<0.5$ which ensures the condition $S<10^{-3}$. Figure~\ref{fig:Fig_2} shows the profile of the mean-field potential from eq.~\ref{U_simple_gaussian} for a typical choice of the parameters, together with the corresponding diagram of the quantum potential from eq.~\ref{V_simple_gaussian}. We emphasize that a bistable quantum potential $\delta V(x)$ is always recovered from such a model of the equilibrium distribution. In particular, increasing the mean-field potential barrier $\Delta U$ does not generate a third minimum at the origin for $\delta V(x)$ as observed with the quartic mean-field potential. \\
     
    With dashed lines in figure~\ref{fig:Fig_2} we have also reported the following parabolic approximations of the two potential functions 
	 \begin{align}
	     U(x)&=\frac{x_0^2}{2\sigma^2}\left( \frac{x}{x_0}\pm 1\right)^2
	     \label{U_min_simple}\\
	     \frac{\delta V(x)}{E_u}
	      &= \frac{x_0^4}{4\sigma^4}\left( \frac{x}{x_0}\pm 1\right)^2-\frac{x_0^2}{2\sigma^2}
	      	      \label{V_min_simple}
	 \end{align}
    which holds far away from the origin when the contribution of one of the two gaussians to the equilibrium distribution in eq. ~\ref{rho_simple_gaussian} is negligible.\\
    
    The potential barrier heights for the $U(x)$ and the $\delta V(x)$ potentials can be obtained in the form:
    \begin{align}
        \Delta U&:=U(0)-U(x_0)=\frac{x_0^2}{2\sigma^2}-\ln{2} \label{Delta_U_simple} \\
        \Delta V&:=\delta V(0)-\delta V(x_0)=E_u\frac{x_0^4}{2 \sigma^4} \label{Delta_V_simple}
    \end{align}
    We stress the inverse squared dependence of the barrier height $\Delta U$ on the control parameter $\sigma/x_0$. For instance the lowering of $\sigma/x_0$ from $1/2$ to $1/5$ increases the barrier from $1.3$ to $11.8$. This effect is even more pronounced in the case of the quantum potential barrier $\Delta V$ because of its quadratic relation to $\Delta U$:
    \begin{equation}
        \frac{\Delta V}{2E_u} = \left(\Delta U +\ln{2} \right)^2
    \end{equation}
    as derived by the comparison of eqs.~\ref{Delta_U_simple} and~\ref{Delta_V_simple}. Such an equation allows the direct conversion of the mean-field barrier $\Delta U$ to the corresponding quantum potential barrier $\Delta V$ within the two Gaussian model.\\
    
\begin{figure}[ht]
        \centering
        \includegraphics[width=0.9\columnwidth]{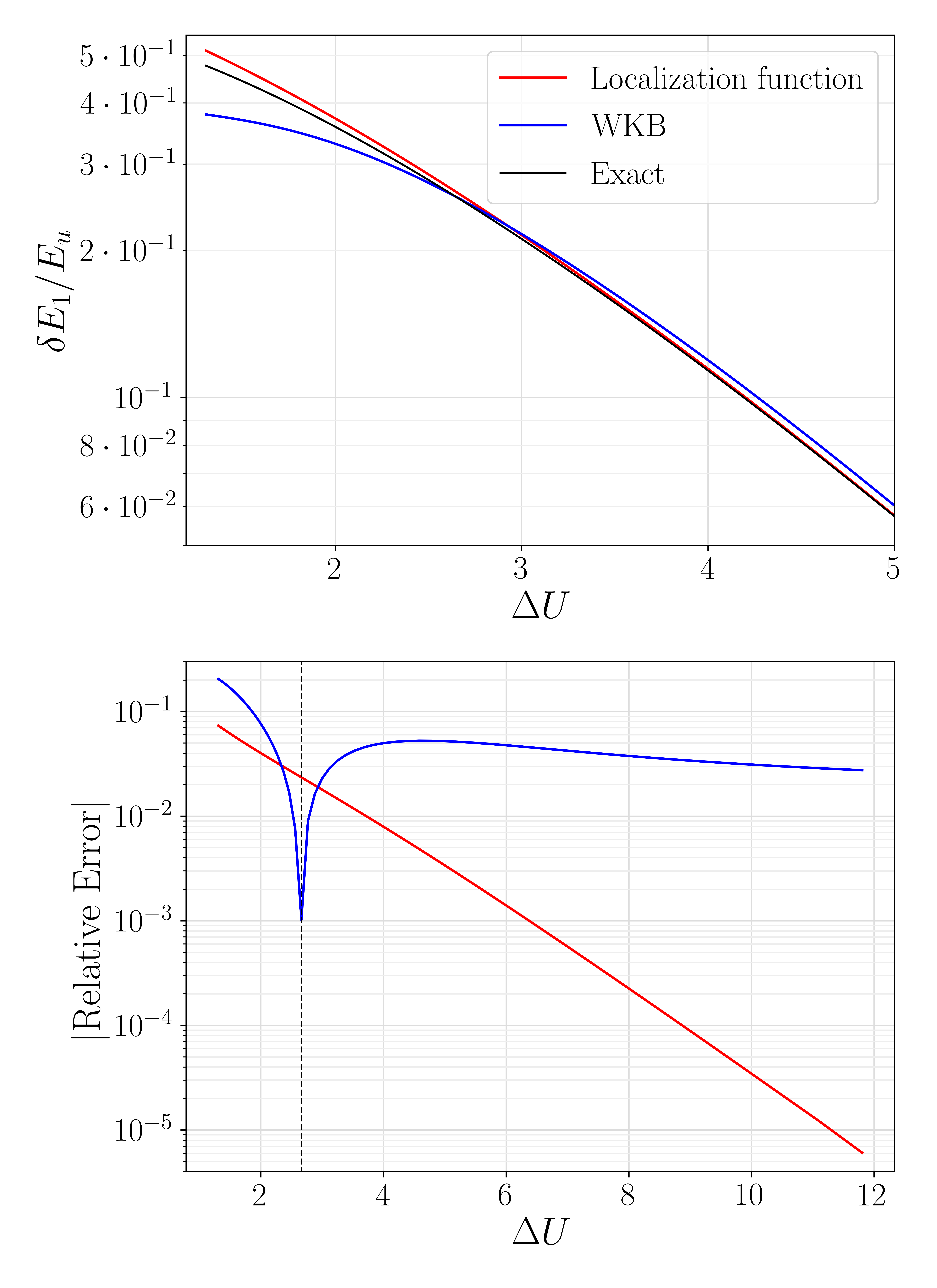}
        \caption{The upper panel shows a comparison between the numerically exact tunneling splitting $\delta E_1$, the localization function integral estimate $\delta E_1^g$ from eq.~\ref{expr_tunneling_splitting} and the WKB value from the Garg~\cite{Garg} formula. The lower panel displays the relative errors of the approximations with respect to the exact tunneling splitting. }
        \label{fig:Fig_3}
    \end{figure}

    With a consistent parameterization of the mean-field potential at hand, the tunneling splitting can now be evaluated according to the localization function approximation $\delta E_1^g$  by numerical calculation of the integrals required by eq.~\ref{expr_tunneling_splitting}. In order to assess its accuracy, depending on the barrier height $\Delta U$, we have compared $\delta E_1^g$ with the exact tunneling splitting $\delta E_1$ obtained by numerical diagonalization of the matrix representation of the Hamiltonian on the basis of a set of Hermite functions, with a careful check on the convergence with respect to the size of the basis set. In figure~\ref{fig:Fig_3} we report, as a function of the mean-field potential barrier $\Delta U$, the exact value $\delta E_1$ (black line) and the localization function approximation $\delta E_1^g$ (red line).
    
    In general, a rather good agreement appears from such a comparison. However, because of the asymptotic origin of the localization function method deriving from the starting approximation in eq.~\ref{FPS_asymptotic_approx}, the error increases by lowering the mean-field potential barrier $\Delta U$. Barriers $\Delta U$ of the order of unity should then represent a situation far from the validity of the asymptotic limit $\Delta U \to \infty$, but still the error continues to be less than $10\%$, and this certifies the quality of the localization function approximation from eq.~\ref{expr_tunneling_splitting}. With barriers $\Delta U$ of few units the deviations of $\delta E_1^g$ from the exact values are already so small that they cannot be perceived by looking at the tunneling splitting estimates displayed in the upper panel of fig.~\ref{fig:Fig_3}. For this reason in the lower panel of fig.~\ref{fig:Fig_3} we have reported the relative error $(\delta E_1^g - \delta E_1)/\delta E_1$ of the approximation  with respect to the exact numerical value $\delta E_1$. The evidence of an exponential convergence, that is of a relative error decreasing exponentially with the barrier $\Delta U$, is recognized from these data and this points out that the localization function methods allows high-quality estimates of the tunneling splitting for intermediate to la large barriers.
    
    For the sake of comparison, we have also considered the WKB estimate of the tunneling splitting in the formulation of Garg~\cite{Garg} as the alternative method supplying an explicit approximation requiring a computational effort comparable with eq.~\ref{expr_tunneling_splitting} of localization function procedure.  The integrals in the WKB approximation have been evaluated numerically by adopting \textit{ad hoc} procedures to counteract the diverging behavior of the integrand near the turning point. The WKB approximation displayed in fig. ~\ref{fig:Fig_3} by blue lines appears much less accurate than the results of the localization function method. It should be noted that at a particular value of the barrier $\Delta U$ corresponding to the vertical dashed line in the lower panel of fig. ~\ref{fig:Fig_3}, the WKB estimate is exact but simply because of the crossing of the two lines describing the barrier $\Delta U$ dependence of the exact and of the WKB tunneling splitting. Notice also that in the same panel we have reported the absolute value of the relative error with a logarithmic scale. These results clearly point out that the convergence of WKB approximation to the exact tunneling splitting with increasing barriers is much slower than in the case of the localization function approximation and that the latter method ensures a gain on the accuracy by orders of magnitude for intermediate to large barriers $\Delta U$.\\
    
\begin{figure}[ht]
        \centering
        \includegraphics[width=0.9\columnwidth]{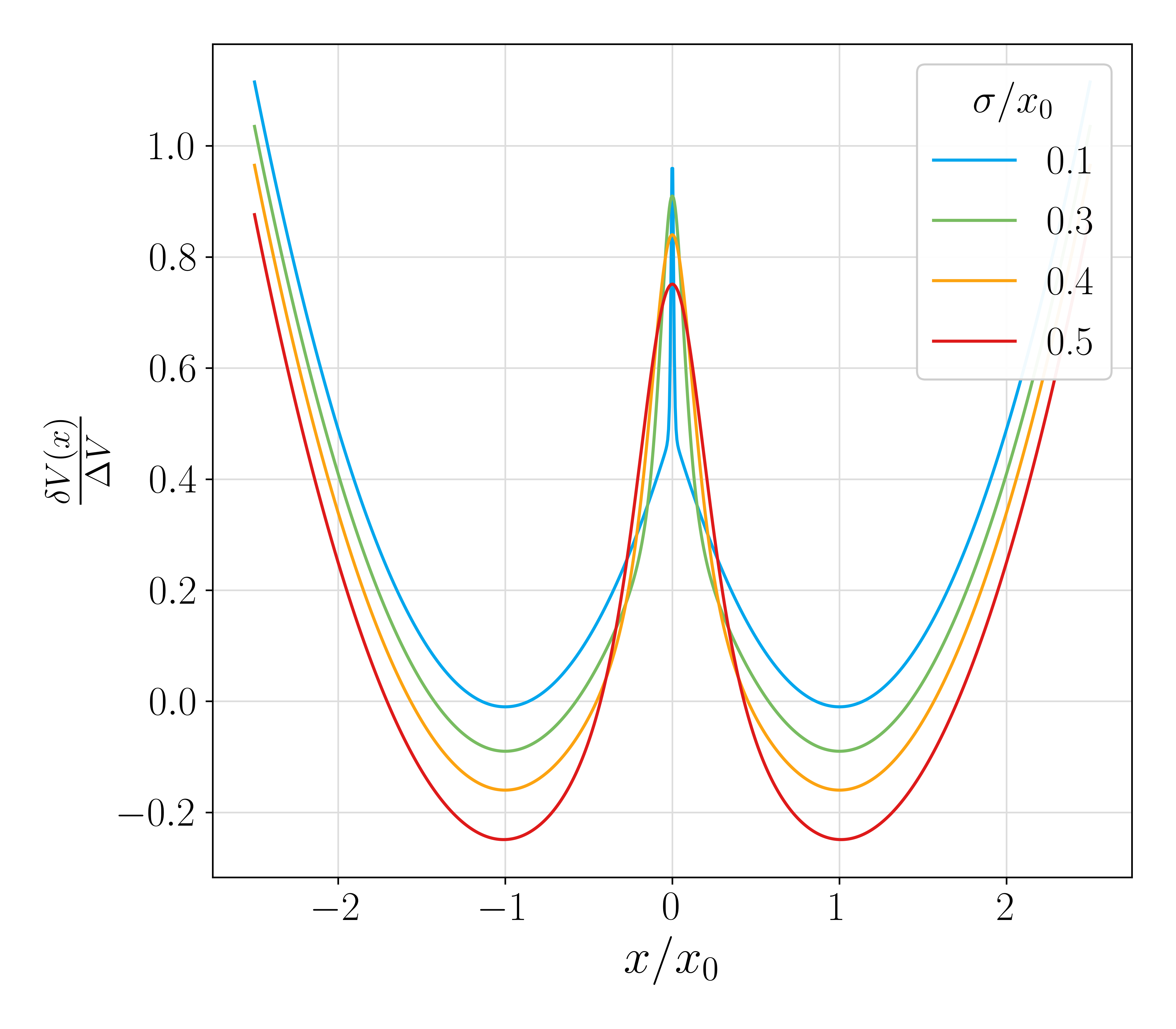}
        \caption{ Effects of the control parameter $\sigma/x_0$ on the quantum potential shape represented as $\delta V(x)$ scaled by its barrier height $\Delta V$. Each profile is labelled by the corresponding value of $\sigma/x_0$.}
        \label{fig:Fig_4}
    \end{figure}

    The previous analysis shows how the tunneling splitting can be easily evaluated according to the two-Gaussian model defined in eq.~\ref{rho_simple_gaussian}. However, one should not conceal the shortcomings of such a simple model in relation to the corresponding quantum potential. This is quite evident by examining the shape of the quantum potential by means of the scaling with respect to the barrier height $\Delta V$, as done in Figure~\ref{fig:Fig_4}. For decreasing values of the control parameter $\sigma/x_0$, the quantum potential develops a cusp-like component at the origin, while one would like to deal with a potential function without any local singular-like component in all the range of independent parameters. Another shortcoming is that this model has only two independent parameters, $x_0$ and $\sigma$, which determine the length scale of the tunneling and the mean-field barrier height $\Delta U$ according to eq.~\ref{Delta_U_simple} or, equivalently, the quantum potential barrier according to eq.~\ref{Delta_V_simple}. This implies that once these features are fixed, there is no more room to choose another important property of the quantum potential: the width of the barrier. In the analysis of the tunneling splitting of specific molecular systems, one needs a parametrized model that is flexible enough to reproduce the most relevant features of the quantum potential. This calls for a three-parameter model in order to take into account, besides the length scale of tunneling and the barrier height, also the barrier width.\\
    
    In order to preserve the methodological benefits of the two-Gaussian model eq.~\ref{rho_simple_gaussian}, we have verified the feasibility of a three-parameter extension by considering an equilibrium distribution with a power dependence on the two-Gaussian linear combination:
    \begin{equation}\label{rho_modulated}
        \rho_{eq}(x):=\frac{N}{\sqrt{8\pi\sigma^2}}\left[ e^{-\frac{(x-x_0)^2}{2\alpha\sigma^2}}+e^{-\frac{(x+x_0)^2}{2\alpha\sigma^2}}\right]^\alpha
    \end{equation}
    in which $N$ is the nearly unitary constant assuring the normalization of the distribution. This extended two-Gaussian model is specified by three parameters: $x_0$ and $\sigma$ as before, and the new parameter $\alpha$ for power modulating the two-Gaussian linear combination. Of course the simpler model of eq.~\ref{rho_simple_gaussian} is recovered for $\alpha=1$. The mean-field potential is defined like in eq.~\ref{U_simple_gaussian}:
    \begin{equation}\label{U_modulated_gaussian}
        \begin{split}
            U(x)&:=-\ln{\left(\frac{\sqrt{8\pi\sigma^2}}{N}\rho_{eq}(x)\right)}=\\
            &=\frac{x_0^2+x^2}{2\sigma^2}-\alpha\ln{\left( e^{\frac{xx_0}{\alpha\sigma^2}}+e^{-\frac{xx_0}{\alpha\sigma^2}}\right)}
        \end{split}
    \end{equation}
    and the corresponding quantum potential is derived  according to eq.~\ref{quantum_vs_thermodynamic_potentials}:
    \begin{equation}\label{V_modulated_gaussian}
        \begin{split}
            \frac{\delta V(x)}{E_u}=\frac{x_0^4}{4\sigma ^4}\left[ \frac{x}{x_0}-\tanh{\left(\frac{xx_0}{\alpha \sigma^2} \right) }\right]^2+\\
            +\frac{x_0^4}{2\alpha \sigma ^4} \text{sech}^2\left(\frac{xx_0}{\alpha \sigma^2}\right)-\frac{x_0^2}{2\sigma ^2}
        \end{split}
    \end{equation}
    Near the Gaussian centers, $x\simeq\pm x_0$, the distribution is well approximated, besides corrections of the order of the superposition coefficient $S:=\exp{\left(-2x_0^2/\alpha\sigma^2\right)}$, by a single Gaussian, $\rho_{eq}(x)\simeq\left(1/\sqrt{8\pi\sigma^2}\right)\exp{\left[-\left(x\pm x_0\right)^2/2\sigma^2\right]}$ independent of the power parameter $\alpha$. Correspondingly the same parabolic expansions in eqs.~\ref{U_min_simple} and~\ref{V_min_simple} are recovered also for the extended two-Gaussian model. The following analysis is confined to the case of well-separated Gaussians to be verified by the smallness of superposition coefficient $S$. With eqs. ~\ref{U_modulated_gaussian} and ~\ref{V_modulated_gaussian} one can evaluate the most important features of the mean-field and of the quantum potentials. In particular, the barrier heights can be computed as:
    \begin{align}
        \Delta U &= \frac{x_0^2}{2\sigma^2} - \alpha\ln{2} \\
        \frac{\Delta V}{E_u} &= \frac{x_0^4}{2\alpha\sigma^4}=\frac{2}{\alpha} (\Delta U +\alpha \ln 2)^2
    \end{align}
    By taking into account that the length unit is arbitrary, one concludes that the interplay between the two control parameters, $\sigma/x_0$ and $\alpha$ allows the independent control of the barrier height and of the width of the quantum potential as desired. The latter parameter $w$ is evaluated, in what follows, as the width of the $\delta V(x)$ quantum potential barriers at half height.

	\begin{table}[ht]
	    \small
        \caption{Parameters for the quantum potentials displayed in Fig.~\ref{fig:Fig_5} at $\Delta V/E_u=30$. Dimensionless units have been employed for the second derivatives $V''(x)$ of the quantum potential by scaling them according to $E_u/x_0^2$. }
        \label{tab:table_1}
        \begin{center}
        \begin{tabular*}{\columnwidth}{@{\extracolsep{\fill}}cccccc}
            \hline
            $\alpha$ & $\sigma/x_0$ & $\Delta U$ & $V''(0)$ & $V''(\pm x_0)$ & $w/x_0$ \\
            \hline
            1.0 & 0.3593 & 3.18 & -2235 & 30 & 0.33\\
            1.5 & 0.3247 & 3.70 & -1124 & 45 & 0.49\\
            2.0 & 0.3021 & 4.09 & -597 & 60 & 0.64\\
            2.5 & 0.2857 & 4.39 & -300 & 75 & 0.76\\
            3.0 & 0.2730 & 4.63 & -115 & 90 & 0.86\\
            \hline
        \end{tabular*}
        \end{center}
    \end{table}
    
\begin{figure}[ht]
        \centering
        \includegraphics[width=0.9\columnwidth]{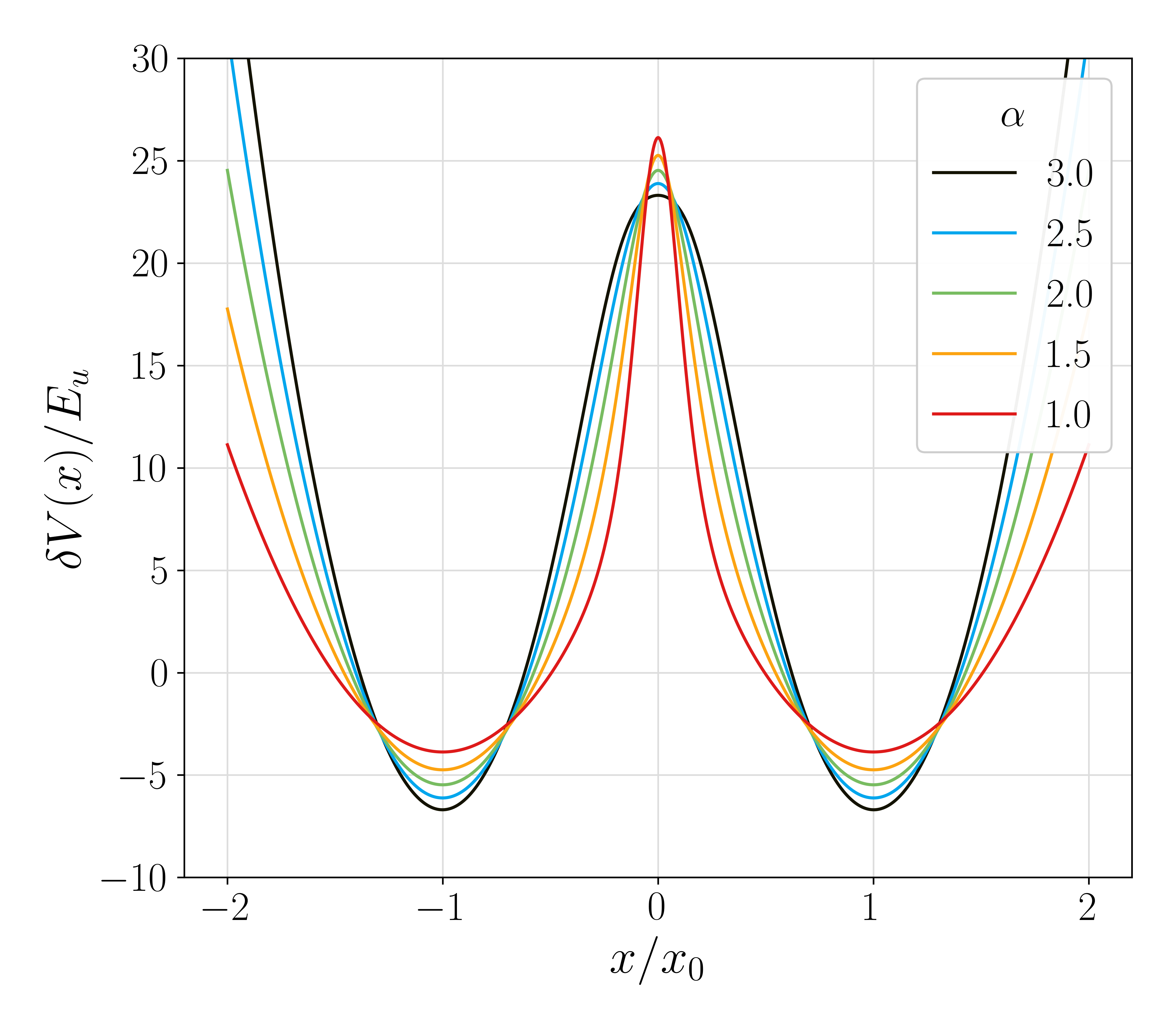}
        \caption{Quantum potentials deriving from the extended two-Gaussian model for the quantum barrier height $\Delta V/E_u=30$. Each profile is labelled by the value of the power coefficient $\alpha$. See Table~\ref{tab:table_1} for the corresponding values of the most significant parameters.}
        \label{fig:Fig_5}
    \end{figure}

    	To characterize the quantum potential generated by the extended two-Gaussian model, we have examined the specific case of a barrier height $\Delta V/E_u = 30$ by sampling the power parameter $\alpha $ with the corresponding values of $\sigma/x_0$ determined by the constraint on the barrier height. A representative set of resulting profiles of the quantum potentials is drawn in Figure~\ref{fig:Fig_5}, while the corresponding values of the most significant parameters are reported in Table~\ref{tab:table_1}. These data show that the increase of the power coefficient $\alpha$ from 1 to 3 decreases by more than an order of magnitude the absolute value of the curvature $V''(0)$ at the saddle point, up to become close to the curvature $V''(\pm x_0)$ of the minima. The visible effect on the shape of the quantum potential is the elimination of the cusp-like behavior at the saddle point to attain a similar parabolic profile, besides the opposite curvature, at the saddle point and the minima. This corresponds to a significant increment of the barrier width. Notice that there is an upper limit on the parameter $\alpha$ ensuring the presence of two minima only in the quantum potential. For example, in the case of fig.~\ref{fig:Fig_5}, a further increase of $\alpha$ would generate positive values for the curvature $V''(0)$, that is the appearance of a third minimum of the quantum potential at $x=0$.\\
    	
    The quantum potential resulting from the extended two-Gaussian distribution  (eq.~\ref{rho_modulated}) is a flexible model which can be used in the study of specific tunneling conditions. In order to define the model, one needs to specify the three main quantum potential features determining the tunneling effect: the distance between the two equivalent minima, the barrier height and the barrier width. From these data, the three independent parameters of the model: $x_0$, $\sigma$ and $\alpha$ can be easily computed.\\
    
    Let us now examine the calculation of the tunneling splitting $\delta E_1$ deriving from this form of the quantum potential. The same  eq.~\ref{expr_tunneling_splitting} for the tunneling splitting according to the localization function method can be applied also to the extended two-Gaussian model. In this case too we confirm the very good accuracy of such an approximation by comparison with the exact numerical solution. In particular, the trend of the relative error is the same as discussed for fig.~\ref{fig:Fig_3}.\\

    \begin{figure}[ht]
        \centering
        \includegraphics[width=0.9\columnwidth]{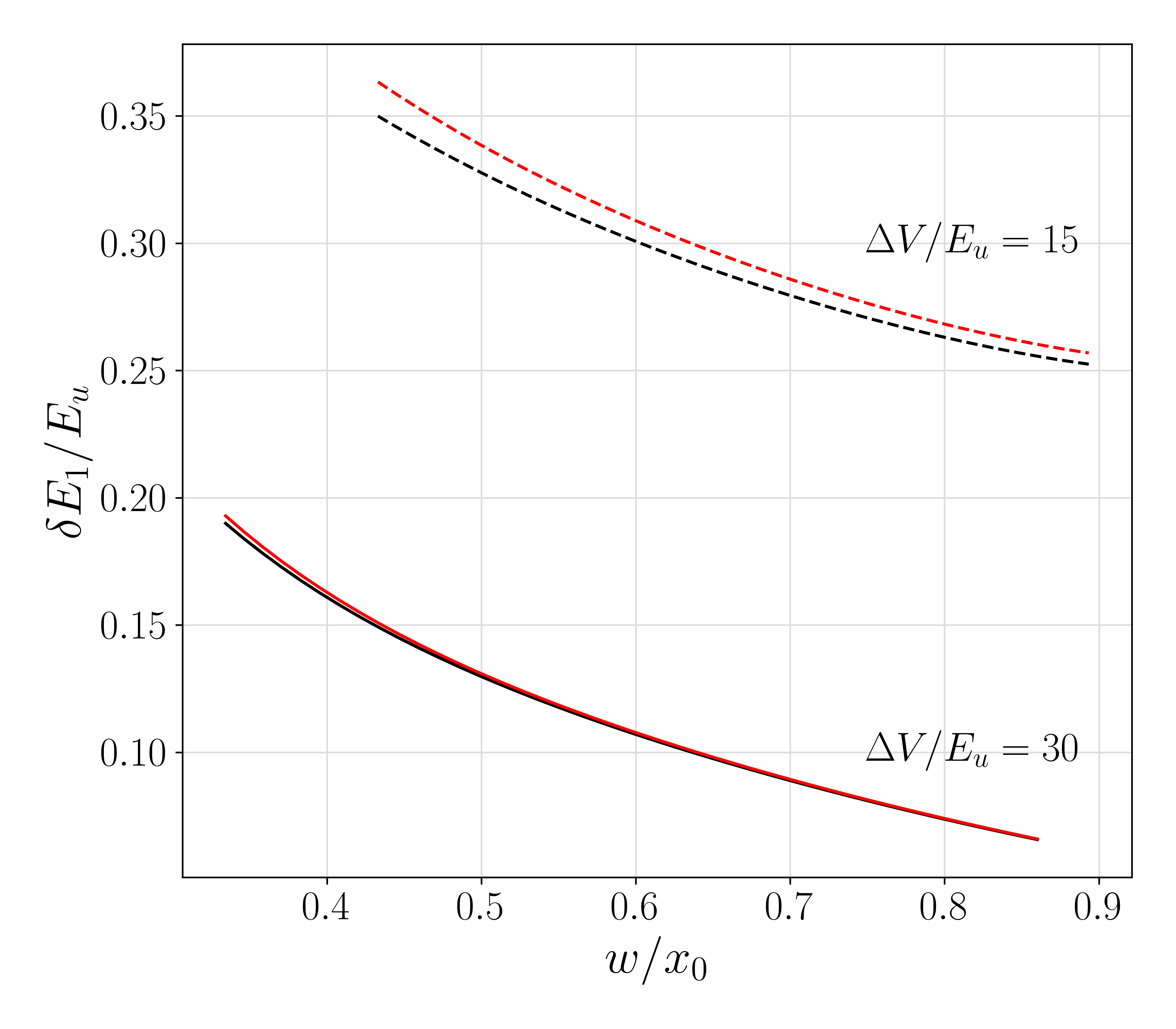}
        \caption{Comparison, for the case of the extended  two-Gaussian model eq.~\ref{rho_modulated},  between the numerically exact tunneling splitting $\delta E_1$ (black lines) with the localization function approximation $\delta E_1^g$ (red lines) as a function of the width $w$ of the quantum barrier for fixed heights (dashed line: $\Delta V/E_u=15$, continuous line: $\Delta V/E_u=30$).}
        \label{fig:Fig_6}
    \end{figure}
  Moreover, such a model allows us to study the effect of the barrier width $w$ at fixed values of the barrier height. In what follows, we discuss the tunneling splitting obtained for values $30$ and $15$ of the barrier height $\Delta V/E_u$. The comparison between exact numerical results (black lines) and the approximation $\delta E_1^g$ resulting from the localization function approach (red lines) is reported in fig.~\ref{fig:Fig_6}. As expected the increase of the barrier width, and correspondingly of the size of coordinate domain representing classically forbidden configurations, produces, for fixed barrier height, a significant decrease of the tunneling splitting. For the $\Delta V/E_u=30$ barrier, the two kinds of results are nearly superimposed and differences can be detected only in the low range of barrier width, that is for the power parameter $\alpha$ close to $1$, which corresponds to lower values of the barrier height $\Delta U$ of the mean-field potential (see table~\ref{tab:table_1}). The differences between the exact and the approximate tunneling splitting are more significant, but still less than 10\%, for the quantum potential barrier $\Delta V/E_u=15$. This is not surprising since the corresponding height of the mean-field barrier is about  $\Delta U \simeq 2$, that is values close to the lower limit for any approximation relying on the gap between the kinetic eigenvalues of the FPS operator and the eigenvalues describing localized relaxation modes~\cite{Moro1995}.
    
\section{Conclusions}\label{sec:conclusions}
    In this paper, we have presented a novel approach to the analysis of the ground state tunneling splitting for one-dimensional bi-stable systems based on the isomorphism between the quantum Hamiltonian and the Fokker-Planck-Smoluchowski operator. We have shown that by applying the localization function method, introduced for the study of activated stochastic processes in the framework of Kramers theory, one obtains an explicit integral approximation for the tunneling splitting which is both asymptotically and variationally justified and therefore capable of accurate predictions in the range of intermediate to large potential barriers. Furthermore, the simple structure of such an approximation allows a direct analysis of the dependence of the tunneling splitting on the main features of the quantum potential like the height and the width of the barrier.\\
    
    The core of the whole procedure is the selection of a parameterized equilibrium distribution that is capable to reproduce the desired shape of the quantum potential. We have shown that this objective can be achieved by employing suitable linear combinations of Gaussian functions.\\
    
    As emphasized in the introduction, the purpose of our work was to explore the potentialities of our method by considering the one-dimensional tunneling splitting problem. The presented results surely support the usefulness of our procedure. The real challenge, however, is represented by its extension to multi-dimensional models that can describe in detail the nuclear dynamics leading to the tunneling splitting of molecules. The known methods for the analysis of activated dynamics described by Fokker-Planck-Smoluchowski equation~\cite{Moro1997, Borkovec, Berezhkovskii2021} can be invoked for this purpose. A more delicate issue arises for the choice of the parameterized equilibrium distribution supporting a multi-dimensional quantum potential of desired shape, but we are confident that suitable linear combinations of multi-dimensional Gaussian functions could provide an effective solution.\\

\section*{Acknowledgements}
We acknowledge the support by the DOR funding scheme of the Dipartimento di Scienze Chimiche, Università di Padova. P. Pravatto is grateful to Fondazione CARIPARO for the financial support (PhD grant).

\printbibliography

\end{document}